\documentclass[11pt]{article}
\begin{document}
\title{Complex Lagrangians and phantom cosmology}
\author{A.A. Andrianov$^{1}$,
F. Cannata$^2$ and A. Y. Kamenshchik$^{2,3}$ }
\date{}
\maketitle
\hspace{-6mm}
$^1$V.A. Fock Department of Theoretical Physics,
Saint Petersburg State University,
198904, S.-Petersburg, Russia\\
$^2$Dipartimento di Fisica and INFN, Via Irnerio 46,40126 Bologna,Italy\\
$^3$L.D. Landau Institute for Theoretical Physics of the Russian
Academy of Sciences, Kosygin str. 2, 119334
Moscow, Russia

\begin{abstract}
Motivated by the generalization of quantum theory
for the case of non-Hermitian
Hamiltonians with PT symmetry, we show how a
classical cosmological
model describes a smooth transition from ordinary dark energy to the phantom one. The model is based
on a classical complex Lagrangian of a scalar field. Specific symmetry properties analogous to PT in
non-Hermitian quantum mechanics lead to purely real equation of motion.
\end{abstract}
\\
\\

\section{Introduction}
Complex (non-Hermitian) Hamiltonians with PT
symmetry have been vigorously investigated in
quantum mechanics and quantum field theory
\cite{Bender}. A possibility of applications to
quantum cosmology have been pointed out in
\cite{non-Hermit1}.  In the present contribution we focus attention on complex classical field theory.
We explore the use of a particular  complex scalar field Lagrangian, which has real solutions of the classical equations of motion.
Thereby we provide a cosmological model describing
in a natural way an
evolution from the Big Bang to the Big Rip involving the transition from
normal matter to phantom matter, crossing smoothly the phantom divide line.

The interest of our approach is related to its focusing on the
intersection between two important fields of research. The basic idea can be introduced qualitatively as follows: given a complex Lagrangian of complex
scalar field, with the complex potential $V(\phi)$,
if $V(\phi)$ is PT symmetric in the sense that
$V(x)$ is PT symmetric in quantum mechanics, e.g.
$V(\phi) \sim \exp(i\alpha\phi)$ with $\alpha$ real,
then Lagrangian becomes real for purely imaginary
$\phi$ and furthermore the kinetic term acquires a
negative sign. In order to avoid any confusion we
stress that our arguments are qualitative and PT
symmetry is not to be intended in a literal way since
 we will deal with spatially homogeneous scalar fields and we have introduced these notations to
refer the reader to the type of potentials and to their symmetry properties we will consider.

The framework in which we work is the general relativity and classical cosmology. We will provide a self-contained introduction to these topics in order that an average
reader can follow the presentation without too many difficulties.

\section{Complex Lagrangians in classical field theory and cosmology}
Let us consider a non-Hermitian (complex) Lagrangian of a
scalar field
\begin{equation}
L = \frac12\partial_{\mu}\phi\partial^{\mu}\phi^* - V(\phi,\phi^*),
\label{Lagrange}
\end{equation}
with the corresponding action, 
\begin{equation}
S(\phi,\phi^*, g) =  \int \, d^4x \sqrt{-|\!|g|\!|} (L + \frac16 R(g)),
\label{action}
\end{equation}
where $|\!|g|\!|$ stands for the determinant of a metric $g^{\mu\nu}$ and $R(g)$ is the scalar curvature term
and the Newton gravitational constant is normalized to $3/8\pi$ to simplify the Friedmann equations further on.

We employ potentials $V(\phi,\phi^*)$
satisfying the invariance condition
\begin{equation}
(V(\phi,\phi^*))^* = V(\phi^*,\phi),
\label{condition}
\end{equation}
while the condition
\begin{equation}
(V(\phi,\phi^*))^* = V(\phi,\phi^*),
\label{condition1}
\end{equation}
is not satisfied.
For example, such potential can have a form
\begin{equation}
V(\phi,\phi^*) = V_1(\phi+\phi^*)V_2(\phi-\phi^*),
\label{factor}
\end{equation}
where $V_1$ and $V_2$ are real functions of their arguments.  
If one defines
\begin{equation}
\phi_1 \equiv  \frac12(\phi + \phi^*),
\label{split1}
\end{equation}
and 
\begin{equation}
\phi_2 \equiv  \frac{1}{2i}(\phi - \phi^*),
\label{split2}
\end{equation}
one can consider potentials of the form
\begin{equation}
V(\phi,\phi^*) = V_0(\phi_1)\exp(i\alpha\phi_2),
\label{factor1}
\end{equation}
where $\alpha$ is a real parameter.  In the last equation one can recognize the link to
the so called $PT$ symmetric potentials.

Here, the functions $\phi_1$ and $\phi_2$ appear as
the real and the imaginary parts of the complex scalar field $\phi$,
however, in what follows, we shall treat them as independent spatially
homogeneous complex
variables depending only on the time parameter $t$.

We shall consider a flat spatially homogeneous Friedmann universe with the metric
\begin{equation}
ds^2 = dt^2 -a^2(t)dl^2,
\label{Fried}
\end{equation}
satisfying the Friedmann equation
\begin{equation}
h^2 = \frac12\dot{\phi}_1^2 + \frac12\dot{\phi}_2^2 +
V_0(\phi_1)\exp(i\alpha\phi_2).
\label{Fried1}
\end{equation}
Here the variable $a(t)$ represents a cosmological radius of the universe and the Hubble variable
$h(t) \equiv \frac{\dot{a}}{a}$ characterizes the velocity of expansion of the universe. The Friedmann
equation (\ref{Fried1}) is nothing but Einstein equation, for the universe filled by scalar fields.

The equations of motion for fields $\phi_1$ and $\phi_2$ have the form
\begin{equation}
\ddot{\phi_1} + 3h\dot{\phi_1} +V_0'(\phi_1)\exp(i\alpha\phi_2) = 0,
\label{motion}
\end{equation}
\begin{equation}
i\ddot{\phi_2}+ 3ih\dot{\phi_2} - \alpha V_0(\phi_1)\exp(i\alpha\phi_2) = 0.
\label{motion1}
\end{equation}
Eqs. (\ref{Fried1}), (\ref{motion}) and (\ref{motion1}) are obtained by variation of the action 
(\ref{action}) with the lagrangian (\ref{Lagrange}) and the potential (\ref{factor1}) with respect to 
the metric, and the scalar field variables $\phi_1$ and $\phi_2$.

\section{Cosmological solutions for accelerated universes}

Let us notice, that the system of equations
(\ref{motion}),(\ref{motion1}),(\ref{Fried1}) can have a solution
where $\phi_1(t)$ is real, while the $\phi_2$ is imaginary, or, in other words
\begin{equation}
\phi_2(t) = -i\xi(t),
\label{imaginary}
\end{equation}
where $\xi(t)$ is a real function. In terms of these
two functions, our system of two real equations can be rewritten as
\begin{equation}
\ddot{\phi_1} + 3\sqrt{\frac12\dot{\phi}_1^2 - \frac12\dot{\xi}^2 +
V_0(\phi_1)\exp(\alpha\xi)}
\dot{\phi_1} +V_0'(\phi_1)\exp(\alpha\xi) = 0,
\label{motion2}
\end{equation}
\begin{equation}
\ddot{\xi} + 3\sqrt{\frac12\dot{\phi}_1^2 - \frac12\dot{\xi}^2 +
V_0(\phi_1)\exp(\alpha\xi)}
\dot{\xi} -\alpha V_0(\phi_1)\exp(\alpha\xi) = 0.
\label{motion3}
\end{equation}

Now, substituting $\phi_2(t)$  from Eq. (\ref{imaginary}) into Eq.
(\ref{Fried1}) we have the following expression for the energy density
\begin{equation}
\varepsilon = h^2 = \frac12\dot{\phi}_1^2 - \frac12\dot{\xi}^2 +
V_0(\phi_1)\exp(\alpha\xi).
\label{energy}
\end{equation}
Thus, we have paved the way to convert an action with a complex potential into the action with real potential and hyperbolic structure of the kinetic term.  
Notice that in the flat Friedmann universe the energy density is always positive, being proportional to
squared Hubble variable which should be real because of reality of geometry.

Let us emphasize that the solution which we are looking for, namely, the imaginary solution 
for $\phi_2$ is the solution which can have sense for the whole system of equations
of Klein-Gordon type (\ref{motion}), (\ref{motion1}) and that of Friedmann (\ref{Fried1}), because 
it makes the energy density real and positive. One can trace some kind of analogy with the non-Hermiatian 
PT symmetric quantum theory with real spectra bounded from below. Our motivation is just to combine the idea of i) providing physical application for ``PT''-symmetric potentials and of ii) building  
of proper framework for conversion of the elliptic structure of the kinetic term for the scalar field 
to the hyperbolic one. In fact, it is crucial to start from a complex Lagrangian with the above mentioned 
symmetry in order that Eqs. (\ref{motion}) and (\ref{motion1}) become real after the  rotation
$\phi_2(t) = -i\xi(t)$. 

We shall see in the section 4 that one can construct such solutions
so that the originally complex Lagrangian becomes real on classical
configurations while one of the scalar components
obtains the ghost (negative) sign of kinetic energy. Thereby we recover a more
conventional phantom matter starting from the complex
matter with normal kinetic energy.

Now, coming back to the Einstein equations we should remember that
the pressure will be equal
\begin{equation}
p = \frac12\dot{\phi}_1^2 - \frac12\dot{\xi}^2 -
V_0(\phi_1)\exp(\alpha\xi).
\label{pressure}
\end{equation}
The pressure in cosmology can be negative. Moreover, the so called dark energy, responsible for recently discovered phenomenon of cosmic acceleration \cite{cosmic},  is characterized by negative pressure such that
$w\equiv p/\varepsilon < -1/3$ \cite{dark}. The value $w =-1$ is nothing but the cosmological constant,
while the dark matter with $w < -1$ was dubbed
as phantom energy \cite{phantom}.
The phantom models have
some unusual properties: to realize them one often uses the phantom scalar
field with the negative sign of kinetic term; in many models the presence
of the phantom dark energy implies the existence of the future Big Rip
cosmological singularity \cite{Rip}.
 A cosmological evolution where dark energy undergoes the transition from $w >-1$ to $w<-1$ implies some particular properties of the corresponding field-theoretical model and is called crossing of phantom divide line \cite{cross}.

From now on we show that the model introduced above is suitable for the description of the phenomenon of the phantom divide line crossing
(for further detail see \cite{us}).

\section{Crossing of the phantom divide line}
It is  easy to see that if $\dot{\phi}_1^2 < \xi^2$ the pressure will be
negative and $p/\varepsilon < -1$,  satisfying the phantom equation
of state. Instead, when $\dot{\phi}_1^2 > \xi^2$, the ratio between
the pressure and energy density exceeds $-1$ and, hence, the
condition
\begin{equation}
\dot{\phi}_1^2 = \dot{\xi}^2
\label{divide}
\end{equation}
corresponds  exactly to the phantom divide line, which can be crossed
dynamically during the evolution of the field components $\phi_1(t)$ and
$\xi(t)$.

We provide  now a simple realization of this idea by
an exactly solvable cosmological model by implementing the
technique for construction of potentials for a given
cosmological evolution \cite{evolution}.
It is convenient to start with a cosmological evolution as given
by the following expression for the Hubble variable:
\begin{equation}
h(t) = \frac{A}{t(t_R-t)}.
\label{evolution}
\end{equation}
The evolution begins at  $t =0$, which represents a standard initial Big Bang
cosmological singularity, and comes to an end in the Big Rip type
singularity at $t = t_R$.  The derivative of the Hubble variable
\begin{equation}
\dot{h} = \frac{A(2t - t_R)}{t^2 (t_R -t)^2}
\label{derivative}
\end{equation}
vanishes at
\begin{equation}
t_P = \frac{t_R}{2}
\label{PDL}
\end{equation}
when the universe crosses the phantom divide line.

Next, we can write down the standard formulae connecting the energy density
and the pressure to the Hubble variable and its time derivative:
\begin{equation}
\frac{\dot{\phi}_1^2}{2} - \frac{\dot{\xi}^2}{2} + V_0(\phi_1)e^{\alpha\xi} =
h^2 = \frac{A^2}{t^2(t_R-t)^2},
\label{Hubble-en}
\end{equation}
\begin{equation}
\frac{\dot{\phi}_1^2}{2} - \frac{\dot{\xi}^2}{2} - V_0(\phi_1)e^{\alpha\xi} =
-\frac23\dot{h}-
h^2 = -\frac{A(4t-2t_R+3A)}{3t^2(t_R-t)^2}.
\label{Hubble-pres}
\end{equation}
The expression for the potential $V_0(\phi_1)$ follows
\begin{equation}
V_0(\phi_1) = \frac{A(2t-t_R+3A)}{3t^2(t_R-t)^2}e^{-\alpha\xi}.
\label{potential}
\end{equation}
The kinetic term satisfies the equation
\begin{equation}
\dot{\phi}_1^2 - \dot{\xi}^2 = -\frac{2A(2t-t_R)}{3t^2(t_R-t)^2}.
\label{kinetic}
\end{equation}

It is convenient to begin the construction with
the solution for $\xi$.
Taking into account the formulae (\ref{evolution}) and (\ref{potential})
Eq. (\ref{motion3}) can be rewritten as
\begin{equation}
\ddot{\xi} + 3\dot{\xi} \frac{A}{t(t_R-t)}
-\frac{\alpha A(2t-t_R+3A)}{3t^2(t_R-t)^2} = 0.
\label{motionxi}
\end{equation}
Introducing a new parameter
\begin{equation}
m \equiv \frac{3A}{t_R},
\label{mdefine}
\end{equation}
Eq. (\ref{motionxi}) looks like
\begin{equation}
\dot{y} + y \frac{mt_R}{t(t_R-t)} -
\frac{\alpha mt_R(2t+t_R(m-1))}{9t^2(t_R-t)^2} = 0,
\label{motionxi1}
\end{equation}
where
\begin{equation}
y \equiv \dot{\xi}.
\label{ydefine}
\end{equation}
The solution of Eq. (\ref{motionxi1})
 is given by
\begin{equation}
y = \frac{\alpha mt_R(t_R-t)^m}{9t^m} \int dt \frac{(2t + (m-1)t_R)t^{m-2}}
{(t_R-t)^{m+2}}.
\label{integral}
\end{equation}

Before considering the concrete values of $m$, notice
that the equation of state parameter $w$ in the vicinity
of the initial Big Bang singularity behaves as
\begin{equation}
w = -1 + \frac{2}{m},
\label{BigBang}
\end{equation}
while approaching the final Big Rip singularity this parameter
behaves as
\begin{equation}
w = -1 - \frac{2}{m}.
\label{BigRip}
\end{equation}
Notice that the range for $w$ does not depend on $\alpha$, depending
only on the value of the parameter $m$, which relates the scales of the
Hubble variable $h$ and of the time of existence of the universe $t_R$.

Remarkably, an integral in the right-hand side of Eq. (\ref{integral})
is calculable analytically
\begin{equation}
\dot{\xi} = \frac{\alpha mt_R}{9t(t_R-t)}
\label{xiint}
\end{equation}
while
\begin{equation}
\xi = \frac{\alpha m}{9} (\log t - \log(t_R-t)).
\label{xiint1}
\end{equation}
From now on the parameter $t$ will be dimensionless. Inclusion of
  characteristic time does not change the structure of the potential because
of its exponential dependence on $\xi$.
Substituting the expression (\ref{xiint}) into Eq. (\ref{kinetic}) one
has
\begin{equation}
\dot{\phi}_1^2 = \frac{mt_R((\alpha^2 m+18)t_R - 36t)}{81t^2(t_R-t)^2}.
\label{phieq}
\end{equation}
For the case $\alpha^2 m = 18$ the function $\phi_1(t)$ can be easily found
from Eq. (\ref{phieq}) and it looks like follows:
\begin{equation}
\phi_1 = \pm \frac{4\sqrt{m}}{3} {\rm Arctanh} \sqrt{\frac{t_R-t}{t_R}}.
\label{phitime}
\end{equation}
One can choose the positive  sign in Eq. (\ref{phitime}) without
loosing the generality.

Inverting Eq. (\ref{phitime}) we obtain the dependence of the time
parameter as a function of $\phi_1$
\begin{equation}
t = \frac{t_R}{\cosh^2\frac{3\phi_1}{4\sqrt{m}}}.
\label{timephi}
\end{equation}
Substituting expressions (\ref{timephi}) and
(\ref{xiint1}) into Eq. (\ref{potential}) we can obtain
the explicit expression for the potential $V_0(\phi_1)$:
\begin{equation}
V_0(\phi_1) = \frac{2\cosh^6\frac{3\phi_1}{4\sqrt{m}}\left(2 + 17
\cosh^2\frac{3\phi_1}{4\sqrt{m}}\right)}{t_R^2}.
\label{potential-result}
\end{equation}
We would like to emphasize that this potential is real and even.
It is interesting that the time dependence of $\phi_1(t)$ could be
found also for an arbitrary value of the parameter $m$, but for
$\alpha^2 m > 18$  this dependence cannot be reversed analytically and,
hence, one cannot
obtain the explicit form of the potential $V_0(\phi_1)$.

\section{Concluding remarks and perspectives}
The phantom model building has involved many different ideas. Here
we have presented a rather simple and natural cosmological toy
model, linked to and inspired by such an intensively developing
branch of quantum mechanics and quantum field theory as the study of
non-Hermitian, but $CPT$ (or $PT$) symmetric models. Notice that
there is an analogy between the manner in which the complexity of
the original Lagrangian with the standard kinetic term is
transformed into phantom-like Langrangian, which is real but has a
negative kinetic energy term and the equivalence between
PT-symmetric quantum Hamiltonians and Hermitian Hamiltonian with
variable effective mass (see e.g. \cite{mass}).

In this paper we have focused essentially on the classical field
theory with a complex potential satisfying the invariance property of Eq. (\ref{condition}). 
Beyond the classical limit one could speculate on how quantum fluctuations $\delta\phi
\equiv \eta_1(x) + i \eta_2(x)$ may preserve the consistency of this  approach.
Assume that the fields of fluctuations respect the initial and final
conditions on a classical solution, $\eta_1(0) = \eta_2(0)= \eta_1(t_R) = \eta_2(t_R)=0$ .
Then the second variation of the action (\ref{action}) reads,
\begin{equation}
S^{(2)}(\eta_1,\eta_2) = \frac12 \int \, d^4x a^3(t) \eta^T \Biggl( -\partial^2_t  - 3 h(t) \partial_t- \hat V^{(2)}(\phi_{1,c}, i\phi_{2,c})\Biggr) \eta,
\label{action2}
\end{equation}
where $\eta^T = (\eta_1(x),\eta_2(x))$ is the transposed field , $\phi_{1,c} \equiv \phi_{1}(t), \, \phi_{2,c} \equiv  - i\xi(t)$ are classical solutions and the matrix $V^{(2)}$ reads,
\begin{equation}
\hat V^{(2)} = \left(\begin{array}{cc}
\partial^2_{\phi_1}V (\phi_{1}, \xi)& i \partial_{\phi_1}\partial_{\xi}V (\phi_{1}, \xi)\\
i \partial_{\phi_1}\partial_{\xi}V (\phi_{1}, \xi)&- \partial^2_{\xi}V (\phi_{1}, \xi)
\end{array}\right).
\label{potential2}
\end{equation}
The quadratic form (\ref{action2}) is symmetric,  with a non-Hermitian but pseudo-Hermitian matrix  $\sigma_3 \hat V^{(2)} \sigma_3 = (\hat V^{(2)})^\dagger$
and the latter fact makes it possible to get real eigenvalues (or pairs of complex conjugated ones) of the energy operator in (\ref{action2}).
In so far as this energy operator is symmetric and has a $2 \times 2$ matrix form, one can diagonalize 
it\footnote{While it is not true in general for complex symmetric matrix in arbitrary dimension as discussed for example in \cite{Gant}, in the case of a quadratic form in Eq.(\ref{action2}), one can straightforwardly show the above statement  by explicit construction of the corresponding complex orthogonal $2\times 2$ matrix.} with an (in general, non-local) orthogonal transformation $\hat O$ so that $\hat O\hat O^T = {\bf I}$.

Because the second variation of potential
$\hat V^{(2)}$ has complex matrix elements, the eigenvectors for a particular real eigenvalue will be also complex. The correct way to perform the variation is: 

first to make an appropriate complex deformation
of the integration contour of variables $\eta_1(x),\eta_2(x) \longrightarrow \tilde\eta_1(x),\tilde\eta_2(x)$  so that the latter complex variables give rise to real ones 
$\eta = \hat O\tilde\eta$; 

next to perform the diagonalization and to end up with a well defined,
real energy operator, hopefully with positive masses. 

To realize this program one has to solve the one-dimensional matrix Schr\"odinger-like equation. To give a concrete idea of what we have in mind we present a solvable  
example of potential (only mildly related to the previous discussion),
$ V = \exp(\sqrt{1 + \alpha^2}\,\phi_1) \exp(i\alpha \phi_2)$.
Its second variation matrix is constant up to an overall factor,
\begin{equation}
\hat V^{(2)} = V \left(\begin{array}{cc}
1 + \alpha^2& i \alpha\sqrt{1 + \alpha^2} \\
i \alpha\sqrt{1 + \alpha^2} &- \alpha^2
\end{array}\right).
\label{potentialex}
\end{equation} 
Its  eigenvalues are $0,1$ and the normalized eigenvectors are \\$e_0^T =(i\alpha,-\sqrt{1 + \alpha^2} ),\, e_1^T =(\sqrt{1 + \alpha^2}, i\alpha )$ so that $e_i^T e_j = \delta_{ij}$ . The diagonalization is realized by the complex orthogonal matrix
\begin{equation}
\hat O = \left(\begin{array}{cc}
i\alpha& - \sqrt{1 + \alpha^2}\\
 \sqrt{1 + \alpha^2} & i \alpha
\end{array}\right).
\label{orthog}
\end{equation} 
The above mentioned contour is given by $\tilde\eta = \hat O^T \eta$ for arbitrary real $\eta$.
Thus, in spite of non-Hermiticity of the matrix of second variation the relevant deformation of the contour makes the action real with positive kinetic terms and masses.

As a last remark we would like to point out that in quantum mechanics, models with CPT symmetry have been 
recently introduced in paper \cite{Znojil} with the difference that the charge operator is a differential 
operator contributing to a definition of a pseudometric operator, whereas in the present approach 
the ``charge'' conjugation is related to a kind of internal degree of freedom.

Finally, let us recapitulate various steps of our approach. \\
1. We relate the possibility of a (de)- phantomization to a  rotation of one of the compenent 
of the scalar field, which gives to the kinetic term a hyperbolic structure.\\
2. The preceding point provides compelling reasons to start from complex PT symmetric Lagrangian.\\
3. We succeed to reduce the problem to two coupled scalar fields: of course one could have started directly from these coupled equations, but in this case, underlying CPT symmetry and CP breaking  
would be hidden \\
We do not claim uniqueness of our approach (see e.g. the recent paper \cite{Chaves}, based on the  superalgebraic approach 
and  using Grassmann vector fields as well as a more general complexification involving the space-time 
coordinates \cite{thooft}), but a definite consequence turns out to be the relation between possible phantomization and CP-breaking.

A.A. was partially supported by RFBR, grant No. 05-02-17477 and by the Programs RNR 2.1.1.1112 and RSS-5538.2006.2. 
A.K. was partially supported by RFBR, grant No. 05-02-17450 and by RSS-1157.2006.2.


\begin{thebibliography}{99}
\bibitem{Bender}
Bender C M  and Boettcher S 1998 {\it Phys. Rev. Lett.} {\bf 80} 5243;
Cannata F, Junker G and Trost J 1998 {\it Phys. Lett.} A {\bf 246 }  219;
Andrianov A A, Cannata F, Dedonder J P and  Ioffe M  V 1999 {\it Int. J. Mod. Phys.}
A  {\bf 14} 2675;
Bender C M, Boettcher S and Meisinger P 1999 {\it J. Math. Phys.}  {\bf 40}
2201;
Bender C M, Milton K.A. and Savage V M 2000 {\it Phys. Rev.} D {\bf 62}  085001;
Bender C M, Brody D C and Jones H F 2004
{\it  Phys. Rev. Lett.} {\bf 93} 251601
\bibitem{non-Hermit1}
Ahmed Z, Bender C M and Berry M V 2005 {\it J. Phys. A: Math. Gen.} {\bf 38}
L627
\bibitem{cosmic}
Riess A et al. 1998
 {\it Astron. J.}  {\bf 116} 1009;
Perlmutter S J et al. 1999
{\it Astroph. J.}  {\bf 517} 565
\bibitem{dark}
Sahni V and Starobinsky A A 2000 {\it Int. J. Mod. Phys.} D {\bf 9} 373
\bibitem{phantom}
Caldwell R R 2002 {\it Phys. Lett.} B {\bf 545} 23;
see the review:  Padmanabhan T 2003  {\it Physics
Reports}  {\bf 380}  235 
\bibitem{Rip}
Starobinsky A A 2000 {\it Grav. Cosmol.} {\bf 6} 157 (astro-ph/9912054);
Caldwell R R, Kamionkowski M and  Weinberg N N
2003 {\it Phys. Rev. Lett.} {\bf 91} 071301
\bibitem{cross}
Feng B, Wang X L and Zhang X M 2005
  {\it Phys. Lett. B} {\bf 607} 35 ;
Vikman A 2005
{\it Phys.  Rev.}  D {\bf 71}  023515;
McInnes B 2005 {\it Nucl. Phys.} B {\bf 718} 55;
Andrianov A A, Cannata F and Kamenshchik A Y 2005 {\it Phys. Rev.} D {\bf 72}
043531;
Aref'eva I Ya, Koshelev A S and Vernov S Yu 2005 {\it Phys. Rev.} D {\bf 72}
064017
\bibitem{us}
Andrianov A A, Cannata F, Kamenshchik A Y,
gr-qc/0512038
\bibitem{evolution}
Starobinsky A A 1998 {\it JETP Lett.}  {\bf 68} 757;
Saini T D, Raychaudhury S,  Sahni V and  Starobinsky A A 2000
{\it Phys. Rev. Lett.} {\bf 85} 1162
\bibitem{mass}
Bagchi B, Quesne C and Roychoudhury R 2005
{\it J. Phys. A: Math. Gen. } {\bf 38} L647;
quant-ph/0511182
\bibitem{Gant}
Gantmacher F R 1959 {\it The theory of matrices} v.2, p. 11 (Chelsea Publishing Company: New York)  
\bibitem{Znojil}
Bagchi B et al 2005 {\it Int. J. Mod. Phys.} A {\bf 20} 7107 
\bibitem{Chaves}
Chaves M and Singleton D 2005 hep-th/0603160
\bibitem{thooft} 't Hooft G and Nobbenhuis S 2006  gr-gc/0602076 
\end{thebibliography}
\end{document}